\documentclass[bibyear]{aa} 

\usepackage{graphicx}
\usepackage{amsmath}
\usepackage{float} 
\usepackage{caption}
\usepackage{xcolor}
\usepackage{soul}

\usepackage{txfonts}
\begin{document}

  \title{Gravity and limb-darkening coefficients for compact stars: DA, DB, and DBA eclipsing   white dwarfs }
 { }

\author{A. Claret \inst{1, 2} \and E. Cukanovaite \inst{3}, K. Burdge \inst{4}, P.-E. Tremblay\inst{3}, S. Parsons \inst{5} \and T. R. Marsh\inst{3}}
   \offprints{A. Claret, e-mail:claret@iaa.es. Tables 1-112 are   available 
in electronic form at the CDS via anonymous ftp or directly from the authors. }
\institute{Instituto de Astrof\'{\i}sica de Andaluc\'{\i}a, CSIC, Apartado 3004,
            18080 Granada, Spain
            \and
            Dept. F\'{\i}sica Te\'{o}rica y del Cosmos, Universidad de Granada, 
            Campus de Fuentenueva s/n,  10871, Granada, Spain\
            \and 
             Department of Physics, University of Warwick, Coventry CV4 7AL, UK 
             \and
            Division  of Physics, Mathematics and Astronomy, California Institute 
            of Technology, Pasadena, CA 91125, USA
             \and
            Department of Physics and Astronomy, University of Sheffield, Sheffield S3 7RH, UK}
            \date{Received; accepted; }

\abstract
   {The distribution of the specific intensity  over the stellar disk is an  
   essential tool for modeling the light curves  in eclipsing binaries,  planetary transits, 
   and stellar diameters through interferometric techniques, line profiles in rotating stars,  
   gravitational microlensing, etc. However, the available theoretical calculations
   are mostly restricted to stars on the main sequence or the giant branch, and very few 
   calculations are available for compact stars.}
   {The main objective of the present work is to extend these 
investigations by computing the gravity and limb-darkening coefficients for white dwarf 
atmosphere models with hydrogen, helium, or mixed compositions (types DA, DB, and DBA).}
   {We computed gravity and limb-darkening coefficients for DA, DB, and DBA 
   white dwarfs atmosphere models, covering  the transmission curves of the Sloan,  
   UBVRI, Kepler, TESS, and Gaia photometric systems. Specific calculations for the HiPERCAM instrument were also carried out. For all calculations of the limb-darkening coefficients 
   we used the least-squares method. Concerning the effects of tidal and rotational 
   distortions, we also computed for the first time the gravity-darkening 
   coefficients $y(\lambda)$ for white dwarfs  using the same models of stellar 
   atmospheres as   in the case of limb-darkening. A more general differential 
   equation was introduced to derive  these quantities, including the 
   partial derivative 
   $\left(\partial{\ln I_o(\lambda)}/{\partial{\ln g}}\right)_{T_{\rm eff}}$.}
   {Six laws were adopted to describe the specific intensity distribution: linear, 
   quadratic, square root, logarithmic, power-2, and a more general one with four 
   coefficients. The computations are presented for the chemical compositions log[H/He] = $-$10.0 (DB), $-$2.0 (DBA) and He/H = 0 (DA), with log g varying between 5.0 and 9.5 and 
   effective temperatures between 3750 K-100\,000 K. For effective temperatures 
   higher than 40\,000 K, the models were also computed adopting nonlocal thermal equilibirum (DA). The  
   adopted  mixing-length parameters are  
    ML2/$\alpha = 0.8$ (DA case) and 1.25 (DB and DBA).     The results are presented in the form of 112 tables. Additional calculations, 
   such as for other photometric systems and/or  different values of log[H/He], $\log g,$ and T$_{\rm eff}$ can be performed upon request.}
   {}

   \keywords{stars: binaries: close; stars: evolution; stars: white dwarfs;
    stars: atmospheres; planetary systems }
   \titlerunning {Limb-darkening coefficients for white dwarfs}
   \maketitle
%

\section{Introduction}

The  distribution of  the specific intensity over the stellar disk is a
very important tool for interpreting the light curves of extrasolar transiting planets and  
double-lined eclipsing binaries, and for studies on the stellar diameters using interferometric 
techniques, line profiles in rotating stars,  gravitational microlensing, etc. 
During the  past few years, this type of information (limb-darkening coefficients, LDC) 
has increased mainly due to the discovery of new exoplanetary systems. 
The LDCs play a very important role in their characterization. However, these coefficients 
are  limited to main-sequence stars and  stars on the giant branch. An exception 
to this rule is the pioneering work of  Gianninas et al. (2013). These authors 
have focused their studies on more compact stars, specifically, white dwarfs 
(hydrogen-rich, DA, model atmospheres). As far as we know, the only study of the gravity-darkening 
exponents (GDE) for compact stars is that by Claret (2015), 
who used  an analytical method to investigate how the temperature is distributed 
over distorted neutron stars.

The main motivation of the present work is to extend the investigations by Gianninas et al. (2013) 
by computing LDC for white dwarf atmosphere models of types DA, DB, and DBA. 
As a complement to the LDC calculations, we present for the first time 
the gravity-darkening coefficients (GDC) calculations for white dwarfs.

In the past decade, the number of known close binary systems containing compact stars has vastly expanded. Many of these systems exhibit photometric variability in the form of eclipses, ellipsoidal modulation, Doppler beaming, etc. Their light curves can therefore serve as a powerful tool for constraining system parameters because light-curve modeling can be used to infer parameters such as the mass ratio, the ratio of component radii to the semi-major axis of the orbit, and inclination. This information can then be combined with information such as spectroscopically measured radial-velocity semi-amplitudes or measured orbital decay due to gravitational wave emission to further constrain the properties of the system. However, a great limitation in modeling binaries involving one or two compact stars has been the absence of accurate limb- and gravity-darkening coefficients for these objects, which determine the surface flux distribution, and thus are crucial in determining the photometric variability of the object. While the effects of limb darkening can be subtle (altering fluxes at the few percent level), modern astronomical cameras such as ULTRACAM (Dhillon et al. 2007) and its successor HiPERCAM (Dhillon et al. 2018) are able to routinely reach this level of precision for white dwarf binaries (e.g., Parsons et al. 2017; Rebassa-Mansergas et al. 2019) and thus accurate limb darkening coefficients are necessary to fully use these facilities for white dwarf science.

Limb darkening affects the shape of the ingress and egress of a white dwarf that is eclipsed. In principle, accurate limb-darkening coefficients permit measuring the binary inclination from just the eclipse, even in well-detached binaries  (e.g., Littlefair et al. 2014), bypassing other expensive or indirect methods for constraining inclinations (Parsons et al. 2017). Moreover, limb-darkening coefficients are invaluable when white dwarfs are modeled in accreting binaries such as cataclysmic variables or AM CVn systems. In these cases, the white dwarf ingress and egress can be affected by accretion processes (e.g., flickering, or light from the accretion disk). Accurate limb-darkening coefficients lift much of the degeneracy when these light curves are modeled (e.g., McAllister et al. 2019). Limb darkening must also be considered when the light curves of transiting planetary debris around a metal-polluted white dwarf such as WD 1145+017 (Hallakoun et al. 2017) is modeled.

The most complete set of limb-darkening coefficients for white dwarfs so far has been presented by Gianninas et al. (2013). These coefficients span a wide range of possible white dwarf parameters, but do not cover the very hot end of the distribution ($T_{\rm{eff}} > 40,000$ K). However, newly discovered systems such as ZTF J1539+5027 (Burdge et al. 2019a) contain one or more hot components that fall within this very hot range. Moreover, the small difference in the filter bandpasses between the Large Synoptic Survey Telescope (LSST) filters used in Gianninas et al. (2013) and the HiPERCAM filters also introduces small systemic errors and thus needs to be taken into account when work at high precision is conducted. Additionally, many of the most compact binary systems involving white dwarfs (e.g., ZTF J1539+5027 and SDSS J0651+2844; Brown et al. 2010, Hermes et al. 2012) are at such a short orbital period that one of the white dwarfs begins to exhibit tidal deformation. This leads to a periodic modulation of the light curve at twice the orbital frequency. This phenomenon strongly depends on the gravity-darkening coefficient and has never been systematically computed for white dwarfs, likely because until recently, no detached binary systems exhibited measurable tidal deformation.

The paper is organized as follows: Section 2 is dedicated to the description of 
the atmosphere models of white dwarfs we adopted here. Section 3 is devoted to the 
LDC calculations, while in Section 4 we present the calculations for the GDC and  
a brief description of the tables.

\section {Description of the DA, DB, and DBA atmosphere models}

For this study we computed three separate grids of 1D DA, DB, and DBA models. Calculations for helium-dominated atmospheres are motivated by the recent discovery of a DBA in a close binary system (Burdge et al. 2019b).

The DA grid spans the same surface gravities, $\log{g}$, as the DA grid of Gianninas et al. (2013): $5.0 \le \log{g} \le 9.5$. We extended the effective temperature range to $3\,750$ K $\le T_{\rm{eff}} \le 100\,000$ K, however. Because DA white dwarfs have nearly  pure hydrogen atmospheres, we used a helium-to-hydrogen ratio, He/H, equal to 0. The atmosphere code described in Gianninas et al. (2013), Bergeron et al. (2011), and Tremblay \& Bergeron (2009) was employed to calculate the DA models in local thermodynamic equilibrium (LTE).  We refer to this code as the Montr\'eal code in the rest of the paper. For $T_{\rm{eff}} \ge 40\,000$ K, we computed two subgrids of DA models with and without non-local thermodynamic equilibrium (NLTE) effects. This was done in order to test the NLTE effects on LDC and GDC. The LTE models computed above $T_{\rm{eff}}=40\,000$ K were computed with the Montr\'eal code, whereas the NLTE models were computed using TLUSTY (Hubeny \& Lanz 1995). For all DA models we used ML2/$\alpha$ = 0.8 (Tremblay et al. 2010). 

Our DB grid spans the same surface gravity range as the DA grid, but with different step sizes (see tables). The effective temperature range spanned by the grid is $11\,000$ K $\le T_{\rm{eff}} \le 40\,000$ K. For this grid we used the hydrogen abundance of $\log{\rm{H/He}} = -10.0$, effectively describing a helium-pure atmosphere. The DBA grid spans the same surface gravity and effective temperature range as for the DB models, but the hydrogen abundance was set to $\log{\rm{H/He}} = -2.0$ in the number of atoms. The majority of helium-dominated atmosphere white dwarfs do show traces of hydrogen, with typical $\log{\rm{H/He}}$ values in the range of $-$4.0 to $-$5.0 (Bergeron et al. 2011). We used the large $\log{\rm{H/He} = -2.0}$ hydrogen abundance in this paper to test the influence of hydrogen on the LDC and GDC. For the DB and DBA models we used ML2/$\alpha$ = 1.25 (Bergeron et al. 2011).

For each grid we calculated the spectral intensity as a function of both wavelength and $\mu$, where $\mu = \cos{\theta}$ and $\theta$ is the angle between the line of sight and the outward surface normal. We used 20 $\mu$ values in all cases, unlike Gianinnas et al. (2013), who used 101 individual $\mu$ values.
On the other hand, 3D convective effects have so far been neglected in the calculations of LDC and GDC coefficients for white dwarfs, but we are planning to apply our techniques to existing 3D DA ($T_{\rm eff} \lesssim$ 14,000 K; Tremblay et al. 2013) and DB(A) (Cukanovaite et al. 2019) grids in a future paper, or upon request to the authors. 

\section{Limb-darkening coefficients for DA, DB, and DBA white dwarfs}

The six LDC laws adopted here are written below. The coefficients  are identified in the respective tables. 

The linear law (Schwarzschild 1906, Russell 1912, Milne 1921)
\begin{eqnarray}
          \frac{I(\mu)}{ I(1)} = 1 - u(1 - \mu),  
\end{eqnarray}

\noindent
the quadratic law (Kopal 1950)

\begin{eqnarray}
        \frac{I(\mu)}{ I(1)} = 1 - a(1 - \mu) - b(1 - \mu)^2,
\end{eqnarray}

\noindent
the square-root law (Díaz-Cordovés\&Giménez 1992),

\begin{eqnarray}
      \frac{I(\mu)}{ I(1)} =  1 - c(1 - \mu) - d(1 - \sqrt{\mu}),
\end{eqnarray}

\noindent
the logarithmic law (Klinglesmith\& Sobieski 1970)

\begin{eqnarray}
\frac{I(\mu)}{ I(1)} =  1 - e(1 - \mu) - f\mu\ln(\mu),
\end{eqnarray}

\noindent
the power-2 law  (Hestroffer 1997) 

\begin{eqnarray}
\frac{I(\mu)}{ I(1)} =  1 - g(1 - \mu^h),
\end{eqnarray}

\noindent
and a four-term law (Claret 2000a) 

\begin{eqnarray}
\frac{I(\mu)}{ I(1)} = 1 - \sum_{k=1}^{4} {a_k} (1 - \mu^{\frac{k}{2}}), 
\end{eqnarray}

\noindent
where  $I(1)$ is the specific intensity at the center of the disk and 
 $u, a, b, c, d, e, f, g, h$, and $a_k$ are the corresponding LDCs.  The 
 specific intensities of the model atmosphere were convolved with  the  transmission 
curves for the Sloan ($u'g'r'i'z'y'$), UBVRI, HiPERCAM, Kepler, TESS, and Gaia passbands. 
 It is not straightforward to apply the least-squares method (LSM)  to the original 20 $\mu$ points because they are not equally spaced. The original distribution can lead to very large weights 
for the $\mu$s near the limb. To avoid this problem,  the LDCs 
were computed for 100 interpolated (equally spaced) $\mu$ points instead of 20, 
as provided originally  by  the Montr\'eal code described in Tremblay \& Bergeron (2009) and Bergeron et al. (2011).  

Two numerical methods 
for computing the LDC are available: LSM,  and flux conservation (FCM). The advantages and disadvantages of each method are exhaustively discussed in Claret (2000a). Based on this 
discussion, we here adoped the LSM to adjust the coefficients of Eqs. 1-6. 
For each law, we computed the merit function, given by

\begin{eqnarray}
{\chi^2}= \sum_{i=1}^{N} \left( {y_i - Y_i}\right)^2
,\end{eqnarray}

\noindent
where $y_i$ is the model intensity at point $i$, $Y_i$ is the fitted function at the same 
point, and $N$ is the number of $\mu$ points. We recall that the biparametric 
laws are only relatively accurate for some regions of the Hertzsprung-Russel (HR) diagram. Therefore, we recommend adopting Eq. 6 because it meets 
the following conditions: a)  it uses a single law that is accurate for the whole HR diagram, 
 b) it is capable 
of reproducing the intensity distribution very well, c) the flux is conserved within a very 
small tolerance, d) it is applicable to different filters as well as to monochromatic values, and  
e) it is  applicable to different chemical compositions,  effective temperatures, 
local gravities, and microturbulent velocities.

The biparametric laws (Eqs. 2-5)  cannot reproduce the 
intensity distribution accurately along the HR diagram, although they do it 
better than the linear law. 
When these laws  are able to represent  the intensity distributions well, 
they are only marginally  accurate in certain intervals of effective 
temperatures and/or log g (see Figs. 4 and 5). 
As a consequence, the user has to divide the HR diagram into areas of laws in order
to use the LDC adequately (see below). Moreover, Eq. 6 provides merit functions  of
the order of 2 magnitudes smaller than any of the laws quoted
previously, mainly in regions close to the limb,  as we describe in the next paragraph.

Figs. 1 and 2 display the relative differences 
between the intensities from the model and those from the  adjustments given by  $[I({\rm{model}})-I({\rm{fit}})]/I({\rm{model}})$ as a function of $\mu$ for two cases: the 
four-term law (Fig. 1) and the power-2 law (Fig. 2). The two figures are on the same scale 
to facilitate comparison. The superiority of the four-term law over the power-2 law is clear 
and the relative differences provided by this approach may be of 
some orders of magnitude smaller than those corresponding to the power-2 law, mainly 
near the limb. When we compare the quality of the adjustments using Eq. 6 with 
those from the other biparametric laws,  the quality of the adjustments provided in 
the first one is still even better than in the case of the power-2 law. 

In order to avoid a biased interpretation of the quality of the adjustments 
illustrated in Figs. 1 and 2, we include in Figs. 3 and 4 the merit function
for all DA models  and all six Sloan passbands adopted in this study.  A direct 
comparison between these two figures shows that the fit quality of the four-term law is always 
far higher than that of the power-2 law 
(almost 2 orders of magnitude). They are only similar in the interval 
 4.4-4.6 in log T$_{\rm eff}$  ,but even so, the four-term law is 
superior. In this interval the specific intensities behave more smoothly, which 
explains the similarity in the behavior of the merit function of these two 
 laws. 

For the biparametric laws alone, we illustrate in Fig. 5 the 
behavior of the square-root law for the same conditions as in Figs 3 
and 4. The $\chi^2$ provided by the power-2 law is similar  
to those given by the  square-root law.  The square-root law provides smaller $\chi^2$ 
than the power-2 law for DB models, showing lower effective temperatures. When a biparametric  law is to be adopted, we therefore 
recommend an inspection of the values of the 
merit function for each law, for each model, and for each passband 
to select the most accurate.

It has been known for a long time that the linear law does not adequately 
describe the distribution of specific intensities when realistic 
stellar atmosphere models are used. Despite this deficiency, this law can be used, 
for example, to compare models with different input physics. In Fig. 6 
we compare the linear LDC for the DB, DA, and DBA models to analyze the 
effect of the hydrogen content on the distribution of specific intensities 
for a constant value of log g = 7.00.  The influence of hydrogen content on the 
slope  is clear: DBA models have higher LDCs up to T$_{\rm eff} \approx$ 20.000 K, while 
the opposite occurs for higher effective temperatures.  
For the DA models, LDCs show an almost linear dependence on effective temperature. Only the DA and DBA models have similar LDCs for a narrow range of effective temperatures (10900 K-15500 K, and this is within the semi-empirical errors).
On the other hand, it is important to 
note that the global differences in the LDC for the three sets of models are larger or of 
the order than  the typical  semi-empirical  
errors that can be derived in the case of high-quality light curves. As a consequence, 
the three types of white dwarfs investigated here might be distinguishable observably using the appropriate LDCs.

As mentioned, the linear approximation is not very adequate, but on the other hand, 
it facilitates visualizing the differences between the LDCs that were computed from different model grids. A comparison of our results with the previous ones from Gianninas et al. (2013) is a 
good consistency check. In Fig. 7 we show this comparison for the $g$ filter. The agreement between the two computations is very good, and the small differences may be due to the modifications that have recently been introduced in the code and numerical precision.

   \begin{figure}
   \includegraphics[height=10.cm,width=8.4cm,angle=0]{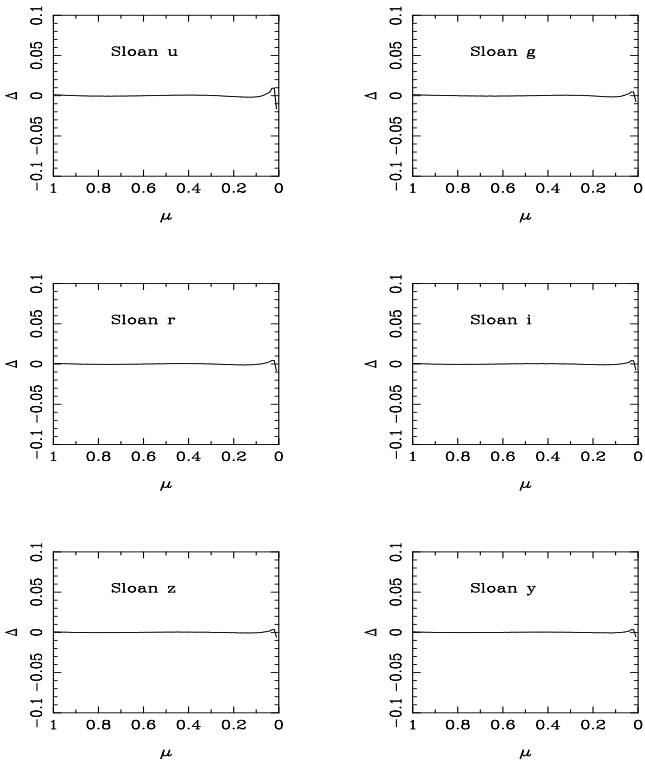}
   \caption{Relative differences $\Delta\equiv$ [I (model) $-$ I (fit)] / I (model) as a function 
   of $\mu$ for the Sloan passbands. Four-term law, log g = 5.5, $T_{\rm eff}$ = 10\,000 K, and DA models (LTE). }
   \end{figure}

   \begin{figure}
   \includegraphics[height=10.cm,width=8.4cm,angle=0]{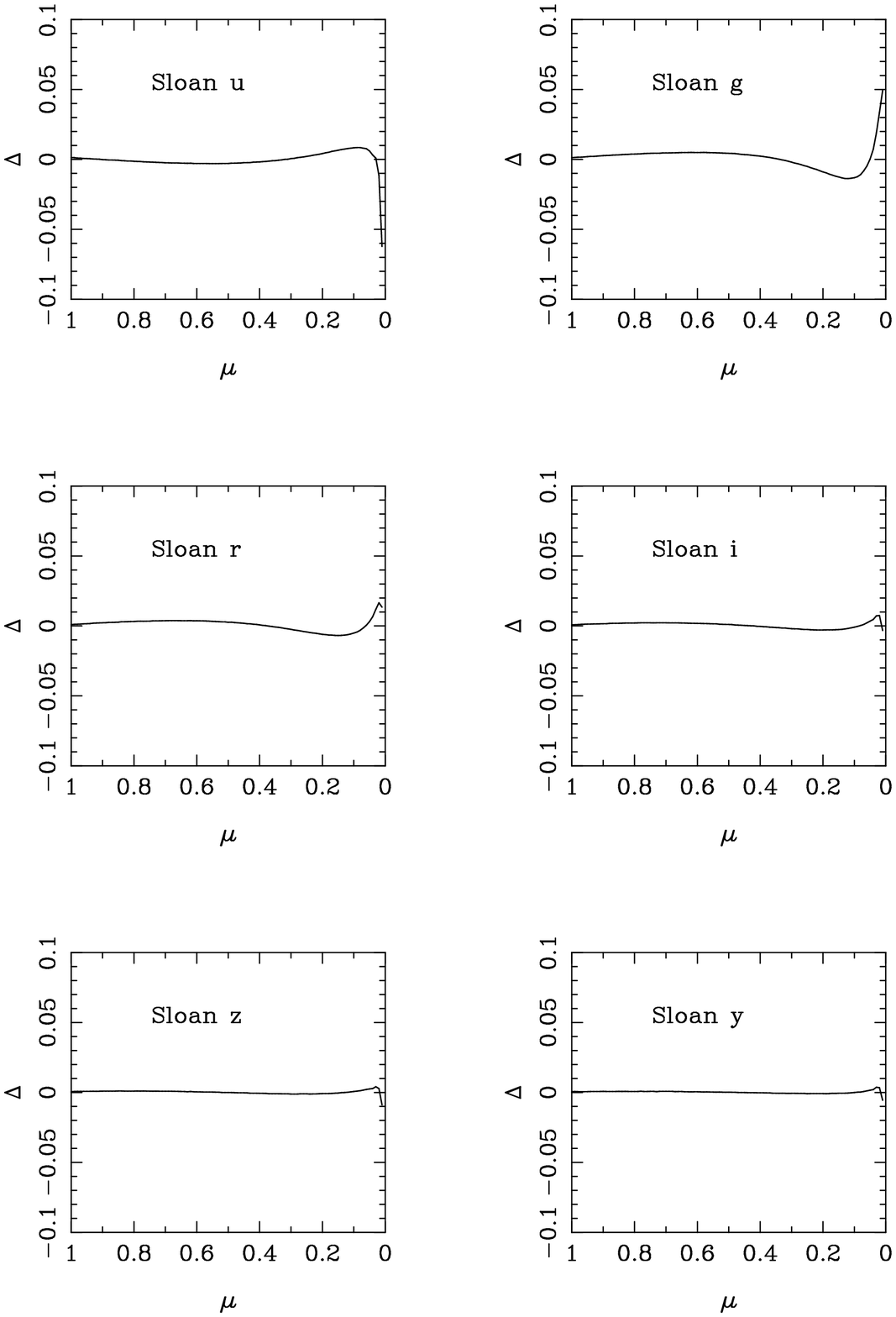}
   \caption{Relative differences $\Delta\equiv$ [I (model) $-$ I (fit)] / I (model) as a function 
   of $\mu$ for the Sloan passbands. Power-2 law, log g = 5.5, $T_{\rm eff}$ = 10\,000 K, and DA models (LTE). }
   \end{figure}

   \begin{figure}
   \includegraphics[height=9.cm,width=9.cm,angle=0]{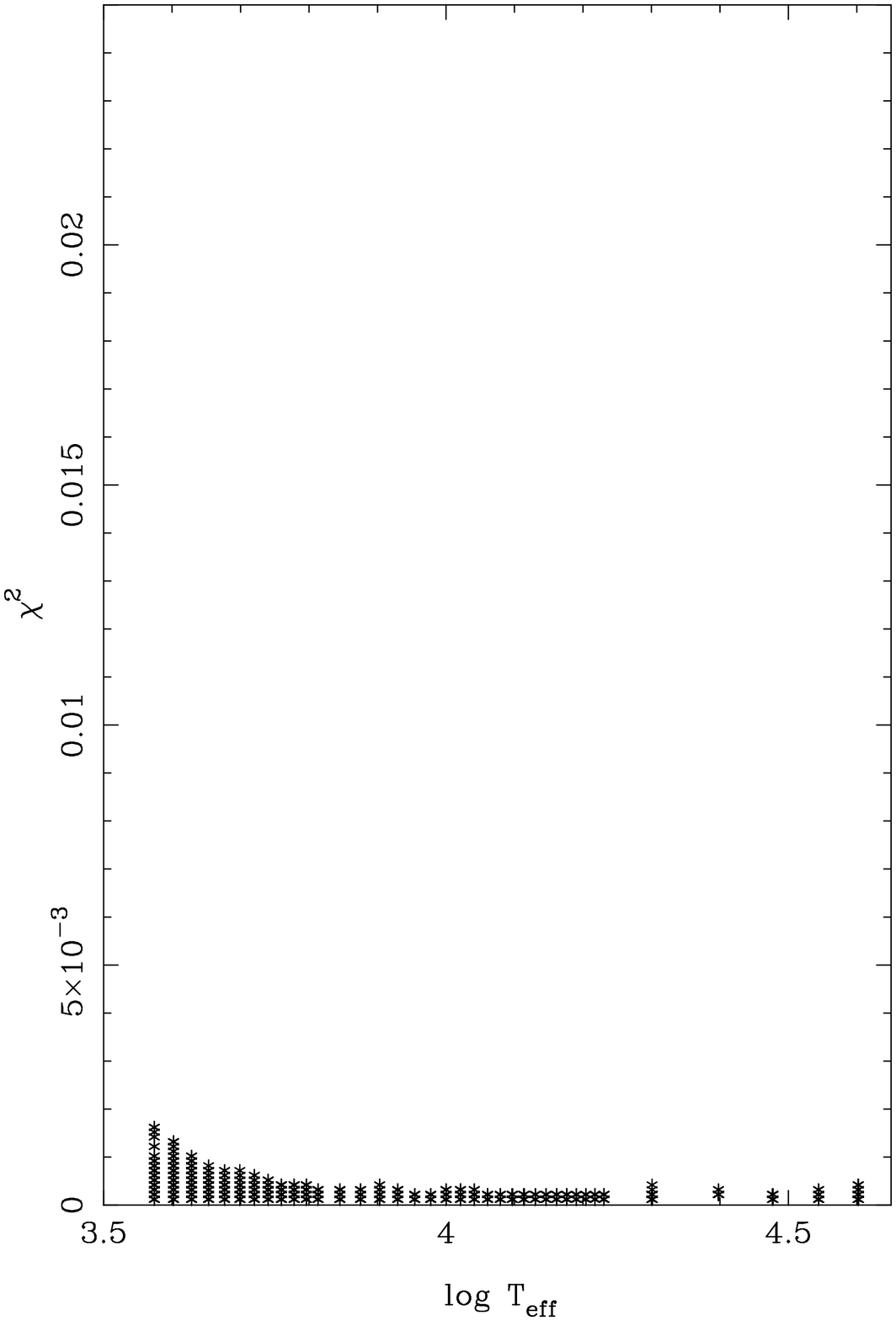}
   \caption{Merit function $\chi^2$ for all models of the sample. Four-term law, 
   all Sloan passbands, and DA models (LTE).}
   \end{figure}

   \begin{figure}
   \includegraphics[height=9.cm,width=9.cm,angle=0]{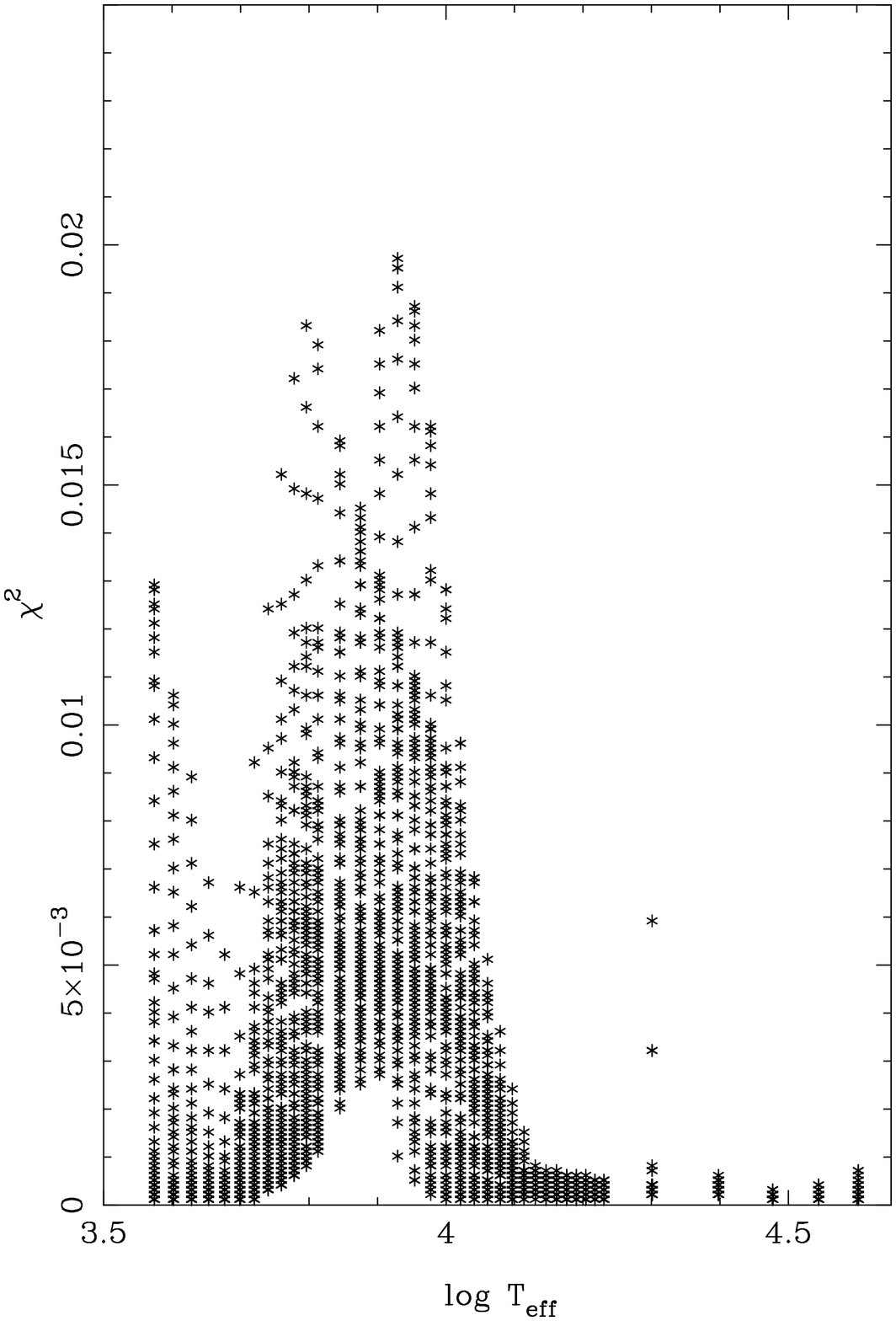}
   \caption{Merit function $\chi^2$ for all models of the sample. Power-2 law, 
   all Sloan passbands, and  DA models (LTE).}
   \end{figure}

   \begin{figure}
        \includegraphics[height=9.cm,width=9.cm,angle=0]{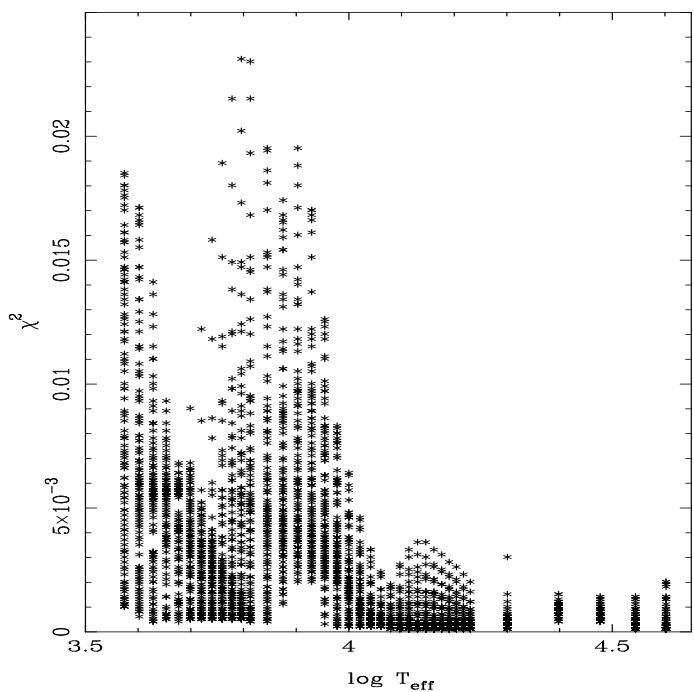}
        \caption{Merit function $\chi^2$ for all models of the sample. Square 
                root-law, all Sloan passbands, and DA models (LTE).}

\end{figure}

   \begin{figure}
        \includegraphics[height=9.cm,width=9.cm,angle=0]{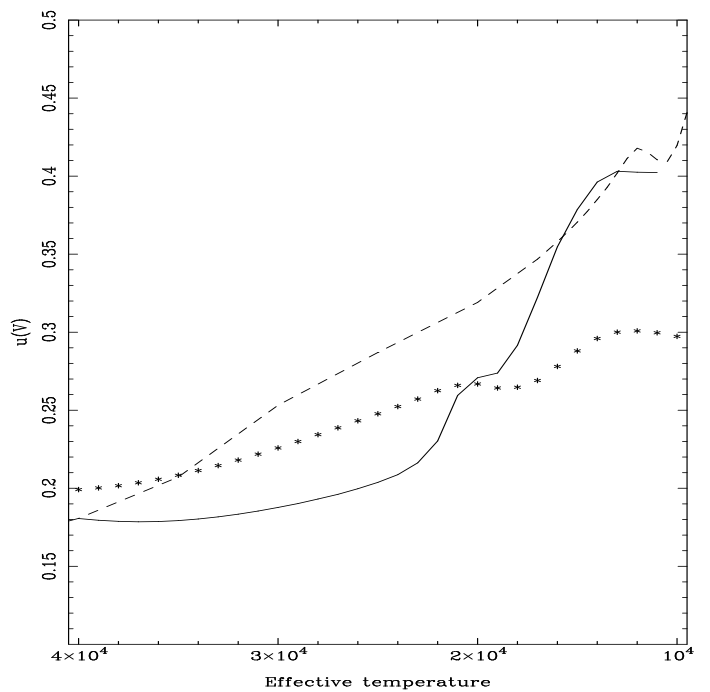}
        \caption{Linear LDC as a function of the effective temperature for the V passband.  Asterisks denote  
                DB models,   DBAs (LTE) are represented by the continuous line, and the 
        dashed line indicates DA models. 
        Log g = 7.0.}
\end{figure}

   \begin{figure}
        \includegraphics[height=9.cm,width=9.cm,angle=0]{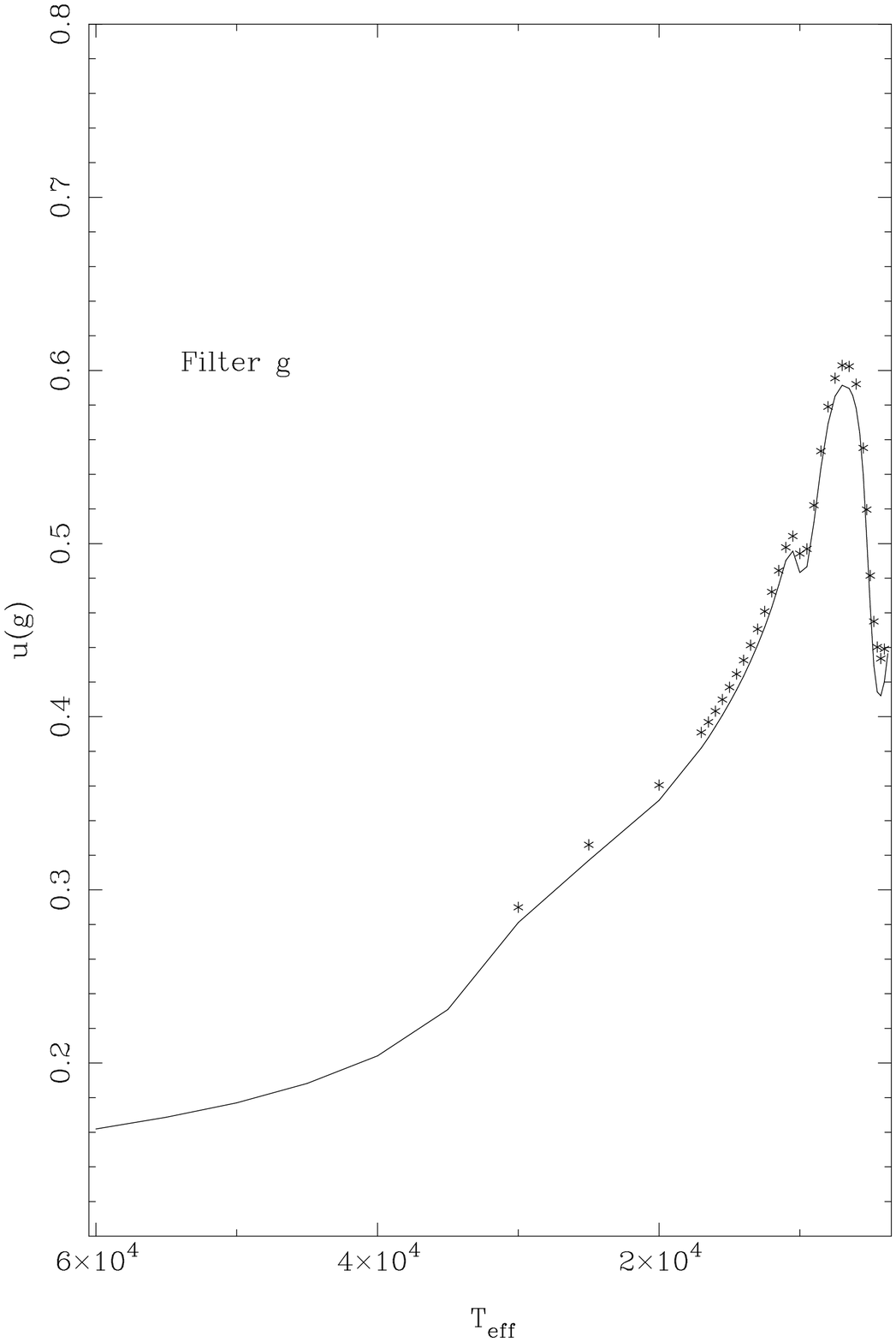}
        \caption{ Comparison between our linear LDCs (continuous line) and those computed by 
                Gianninas et al. (2013), which are denoted  by asterisks.  DA models, log g = 6.0. }
\end{figure}

\section {Calculating the gravity-darkening coefficients}
 
As is well known,  rotation and/or tides  distort  the 
shapes of stars.  
Such distortions have been investigated by Kopal (1959), who used spherical harmonics  
 to describe them. The studies are in particular important in close binary systems or in planetary systems 
where both mechanisms 
act simultaneously:  tides tend to elongate the star, while rotation 
tends to  flatten it at the poles. The corresponding deviations from a sphere 
are functions of the rotational rates and of the mass ratio $q$. A 
mathematical summary will help us understand the results obtained by Kopal. 
Consider the total potential of a rotating star. We assume that the 
angle between the normal and the radius vector is small and that only the radial 
component of the derivative is considered to compute the local gravity. Thus, we have

\begin{eqnarray}{{g-g_o}\over{g_o}}={\sum_j}\left(1-{5\over{\Delta_j}}\right)
\left({r\over{a}}-1\right)  
,\end{eqnarray}

\noindent
where $g_o$ is the local gravity (taken as reference),  $\Delta_{j} = 1 + 2 k_j$ 
and $k_j$ is the apsidal motion constant 
of order $j$. For $j$ = 2, the radius of an equipotential $r$ can be written as

\begin{eqnarray}r = a\left(1 - f_2 P_2(\theta, \phi)\right)  
,\end{eqnarray}

\noindent
where $f_2$ is given by

\begin{eqnarray}f_2 = {5\omega^2a^3\over{3GM_{\psi}(2+\eta_2)}}.  
\end{eqnarray}

In the above equation $\omega$ is the angular velocity, P$_2(\theta, \phi)$ 
is the second surface harmonic, $a$ is the mean radius of the level surface, $\eta_2$ is 
the logarithmic derivative of the spherical harmonic defined through
Radau's equation, and $M_{\psi}$ is the mass enclosed by an equipotential. 
The importance of Eq. 10 is that it establishes a dependence of the GDE 
 on the shape of the distorted stellar configuration, 
on its internal structure, and also on the details of the rotation law. 
More details on these calculations have been presented by Claret (2000b) .

From 
the physical point of view,  more than the geometrical changes must be 
considered. von Zeipel (1924) showed that for configurations in pure radiative 
equilibrium (pseudo-barotrope), the emerging flux is not constant on the surface 
of a distorted star and depends on the value of the local gravity:

\begin{eqnarray}
 {F} = -{4 a c T^{3}\over{3 \kappa \rho}}{dT\over{d\Phi}} {g^{\beta_{1}}}
\end{eqnarray}

\noindent
or

\begin{eqnarray}
 {{T_{\rm eff}}^4} \propto {g^{\beta_1}} 
,\end{eqnarray}

\noindent
where $g$ is the local gravity, $\Phi$ is the total potential,  T is the local 
temperature, $\kappa$ is the opacity, $\rho$ is the local density, $a$ is 
the radiation pressure constant, $c$ the velocity of light in vacuum, T$_{\rm eff}$ is 
the effective temperature, and $\beta_1$=1.0 is the GDE, 
a bolometric quantity. 

However,  significant deviations 
from  von Zeipel's theorem have been found when the GDEs are calculated at 
the upper layers  of a distorted star (Claret 2012). 
This is a consequence of using  a  transfer equation that is more elaborate 
than the diffusion approach that was adopted by von Zeipel. This theorem 
is therefore not strictly valid at lower optical depths.  For details on the calculations  
of  $\beta_1$, recent results and different methods, see for example, Claret (2012, 2016) and 
Zorec et al. (2017). The non-validity of von Zeipel's theorem also extends to compact stars. 
As previously commented in the Introduction, Claret (2015) used a perturbation theory and  
derived an equation for the GDE for neutron stars as a function of the
rotation law, of the colatitude, and of the logarithmic derivatives of the opacity. 
This equation predicts significant deviations from the von Zeipel’s theorem 
for differentially rotating neutron stars.  

Concerning the passband gravity-darkening  coefficients $y(\lambda)$ (GDC), 
 Kopal (1959) 
assumed a simple but ingenious approach six decades ago in the absence of reliable atmosphere models: the distorted configurations radiate like 
a blackbody.  By expanding  the ratio  between the monochromatic and total radiation 
in a Taylor series, he  derived the GDC as a function of the temperature (effective) 
and of the wavelength.  Of course,  the blackbody radiation is not a good 
approximation, and more elaborate atmosphere models are needed to permit a consistent  
light-curve analysis. Later, Martynov (1973) refined the calculation of the GDC 
introducing the following equation:

\begin{eqnarray} 
y(\lambda, T_{\rm eff}, \log [H/He], \log g) = \frac{1}{4} 
\left(\frac{\partial\ln I_o(\lambda)} {\partial\ln T_{\rm eff}}\right)_{g},     
\end{eqnarray}

\noindent
where $\lambda$ is the wavelength, log [H/He] is the content of hydrogen relative to the content of helium, $I_o(\lambda)$ is the specific 
intensity at a given 
wavelength  at the center of the stellar disk, and the  subscript $g$ indicates 
a  derivative  at constant surface gravity. At that time and for many years, 
the blackbody approach was used 
to calculate the partial derivative. We also note that Eq. 13 
also presents some simplifications. The factor 1/4 indicates that 
in its derivation, $\beta_1$ was assumed to be 1.0 for any effective temperature. 
However, as we previously commented, $\beta_1$ is a function of the physical conditions in the upper layers and can be different from 1.0, mainly 
for stars presenting  convective envelopes (see below). 

Claret \& Bloemen (2011) considered the effects of the partial derivative 
$\left(\partial{\ln I_o(\lambda)}/{\partial{\ln g}}\right)_{T_{\rm eff}}$ and of more realistic 
stellar atmosphere models (PHOENIX and ATLAS) 
in calculating the GDC,

\begin{eqnarray}
\lefteqn{y(\lambda, T_{\rm eff }, \log [H/He], \log g, ) =} \nonumber \\ 
\lefteqn{\left(\frac{d\ln T_{\rm eff }}{{d\ln g}}\right)
\left(\frac{\partial{\ln I_o(\lambda)}}{{\partial{\ln T_{\rm eff }}}}\right)_{g}
 + \left(\frac{\partial{\ln I_o(\lambda)}}
{\partial{\ln g}}\right)_{T_{\rm eff}}.}
\end{eqnarray}

Following the arguments presented above, 
the term $\left(\frac{d\ln T_{\rm eff }}{{d\ln g}}\right)$ can 
be written as $\beta_1/4,$ and finally we have

\begin{eqnarray}
\lefteqn{y(\lambda, T_{\rm eff }, \log [H/He], \log g) =} \nonumber \\ \lefteqn{\left(\frac{\beta_1}{4}\right)
\left(\frac{\partial{\ln I_o(\lambda)}}{{\partial{\ln T_{\rm eff }}}}\right)_{g}
 + \left(\frac{\partial{\ln I_o(\lambda)}}
{\partial{\ln g}}\right)_{T_{\rm eff}}.}
\end{eqnarray}

We used the same DA, DB, and DBA models as described in Section 2 to compute the GDC 
for all passbands considered here. 
The corresponding GDC tables consist of two lines per model: the first line refers to 
the first term of Eq. 15 considering $\beta_1$ = 1.0, and the second line refers 
to the second term of this equation. We chose to separate 
the two contributions so that the value of 
$\beta_1$ can be adjusted during the synthesis of the light curves. In this case, the first term only needs to be multiplied by the value of $\beta_1$  to be iterated.

   \begin{figure}
   \includegraphics[height=10.cm,width=8.4cm,angle=0]{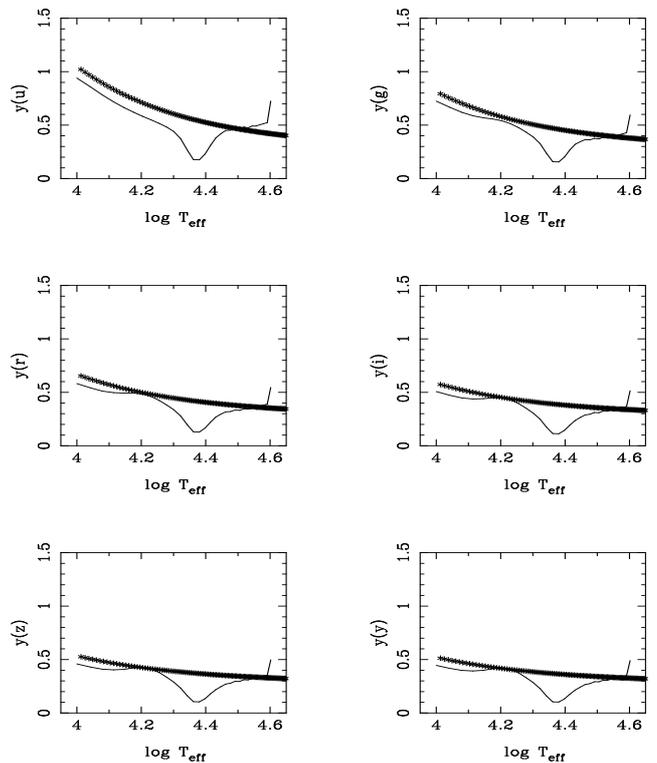}
   \caption{Theoretical gravity-darkening coefficients for the {\sc Sloan} 
   passbands (DB models). The continuous line represents the calculations    for  log $g$ = 8.0, and asterisks denote the calculations by adopting the blackbody approach. }
   \end{figure}

\begin{figure}
        \includegraphics[height=10.cm,width=8.4cm,angle=0]{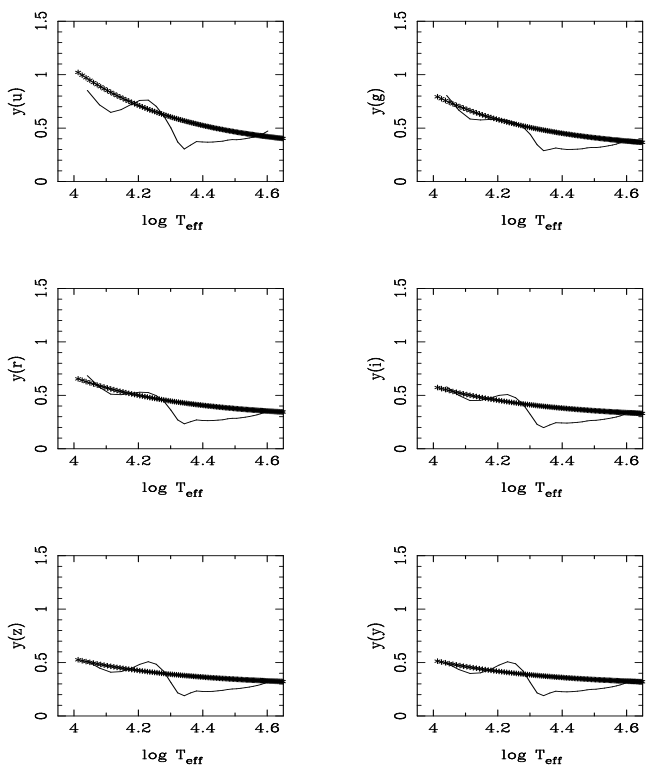}
        \caption{Effects of hydrogen content on the GDC (DBA  models). The captions are the same as for Fig. 8.  }
\end{figure}

\begin{figure}
        \includegraphics[height=10.cm,width=8.4cm,angle=0]{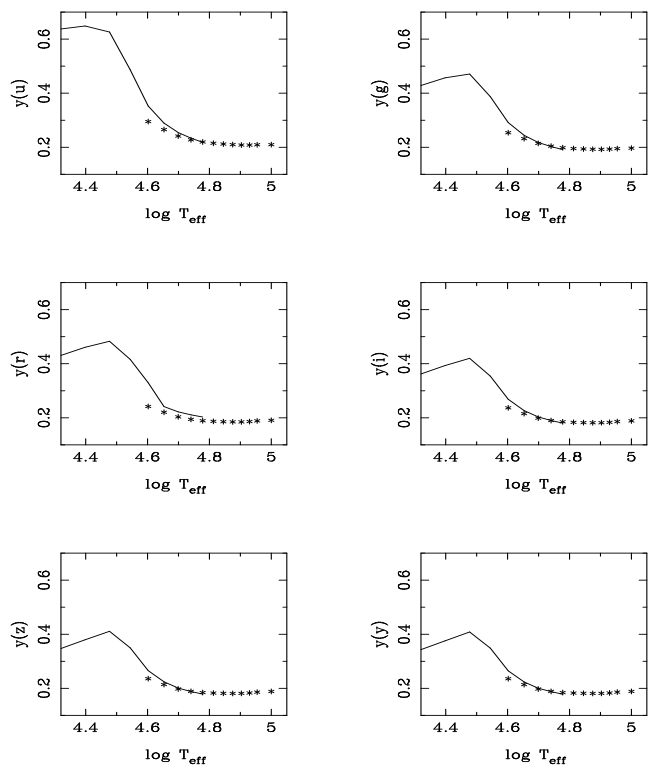}
        \caption{Effects of NLTE on the GDC (DA models). The continuous line indicates LTE models, and asterisks denote NLTE models.  Sloan passbands. Log g = 8.25.}
\end{figure}

Figures 8 and 9   (DB and DBA, respectively) show the comparison between the GDC given by Eq. 15 (continuous line) 
and those calculated using the blackbody approach (asterisks). As anticipated, the blackbody radiation 
is not a good approximation. This is  particularly valid  for  
  passbands with short  effective wavelengths. Similar  effects have been detected 
 in main-sequence and giant stars (see Fig. 4 in Claret 2003).  The influence of the second term in the GDC in Eq. 15 is not  important, but not negligible (see the second lines of the corresponding tables).  We also  detected a decrease   
 in GDC for all Sloan passbands around  20000 K for the DB and DBA models.  This feature might be
connected to the onset of He II convection (Cukanovaite et al. 2019).  

As explained in Sect. 2, we also computed DA NLTE models. GDC calculations can serve as an indicator (although not the definitive) of the effective temperature limit for which it is necessary to adopt NLTE. Fig. 10 shows the comparison between LTE (continuous line) and NLTE (asterisks) models. For all Sloan filters, this effective temperature limit is at about 40\,000 K, in good agreement with previous estimates (Gianninas et al. 2013). This effective temperature is practically independent of local gravity. The systematic and small offsets that we show in Fig. 10 at lower effective temperatures  are probably due to the use of different codes for LTE and NLTE. Therefore we can conclude that the effect of considering NLTE is practically negligible in GDC calculations. A discussion for the onset of NLTE effects can be found in Liebert, Bergeron,  \& Holberg (2005). 

Because 1D convection is limited, the theoretical GDC (DA models) for $T_{\rm eff}$ = 5000 K show some discontinuities at  the shortest wavelengths (passbands $u$ and $U$). We have preferred to keep these discontinuities in the respective tables and leave the task of smoothing them to the user according to their needs.

\begin{acknowledgements} 
 We thank the anonymous referee for helpful comments that
have improved the manuscript.   
 The Spanish MEC (AYA2015-71718-R and
ESP2017-87676-C5-2-R) is gratefully acknowledged for its 
support during the development of this work. A.C. also 
acknowledges financial support from the State Agency for 
Research of the Spanish MCIU through the “Center of 
Excellence Severo Ochoa” award for the Instituto de 
Astrofísica de Andalucía (SEV-2017-0709). The research leading to these results has received funding from the European Research 
Council under the European Union's Horizon 2020 research and 
innovation programme n. 677706 (WD3D). This research has made 
use of the SIMBAD database, operated at the CDS, Strasbourg, 
France, and of NASA's Astrophysics Data System Abstract Service.
\end{acknowledgements}

{}

\begin{appendix}

\section{Brief description of tables A1, A2, and A3}

In order to provide users with the complementary tools for 
the synthesis of light curves, we also computed the Doppler 
beaming for each photometric system.   
 The Doppler beaming factors adopting 
the BB approach were computed according to Eq. 3  (parameter $\alpha$) in 
Loeb\&Gaudi (2003). Tables A1, A2, and A3 summarize the type 
of data available as well as the central intensities for each 
photometric system  in ergs/cm$^2$/s/Hz/ster, and the Doppler 
beaming for the BB approach. For more details, see the ReadMe file 
on the CDS.

        \renewcommand{\tablename}{Table }       
\begin{table*}  

        \caption{Gravity and limb-darkening coefficients for the Sloan and UBVRI photometric systems}
        \begin{flushleft}
                \begin{tabular}{lcccccclc}                         
                        \hline                         
                        Name    & Source   &  range T$_{\rm eff}$ & range log $g$ & log [H/He] & Filters & Fit/equation   \\ 
                        \hline   
Table1  &{\sc DB}  &10000 K-40000 K & 5.5-9.5&  -10.0 &{u'g'r'i'z'y'} & LSM/Eq. 1 \\
Table2  &{\sc DB}  &10000 K-40000 K & 5.5-9.5&  -10.0 &{u'g'r'i'z'y'} & LSM/Eq. 2 \\
Table3  &{\sc DB}  &10000 K-40000 K & 5.5-9.5&  -10.0 &{u'g'r'i'z'y'} & LSM/Eq. 3 \\
Table4  &{\sc DB}  &10000 K-40000 K & 5.5-9.5&  -10.0 &{u'g'r'i'z'y'} & LSM/Eq. 4 \\
Table5  &{\sc DB}  &10000 K-40000 K & 5.5-9.5&  -10.0 &{u'g'r'i'z'y'} & LSM/Eq. 5 \\
Table6  &{\sc DB}  &10000 K-40000 K & 5.5-9.5&  -10.0 &{u'g'r'i'z'y'} & LSM/Eq. 6 \\
Table7  &{\sc DB}  &10000 K-40000 K & 5.5-9.5&  -10.0 &{u'g'r'i'z'y'} & GDC $y(u'g'r'i'z'y')$\\
Table8  &{\sc DB}  &10000 K-40000 K & 5.5-9.5&  -10.0 &{UBVRI}  & LSM/Eq. 1 \\
Table9  &{\sc DB}  &10000 K-40000 K & 5.5-9.5&  -10.0 &{UBVRI}  & LSM/Eq. 2 \\
Table10 &{\sc DB}  &10000 K-40000 K & 5.5-9.5&  -10.0 &{UBVRI}  & LSM/Eq. 3 \\
Table11 &{\sc DB}  &10000 K-40000 K & 5.5-9.5&  -10.0 &{UBVRI}  & LSM/Eq. 4 \\
Table12 &{\sc DB}  &10000 K-40000 K & 5.5-9.5&  -10.0 &{UBVRI}  & LSM/Eq. 5 \\
Table13 &{\sc DB}  &10000 K-40000 K & 5.5-9.5&  -10.0 &{UBVRI}  & LSM/Eq. 6 \\
Table14 &{\sc DB}  &10000 K-40000 K & 5.5-9.5&  -10.0 &{UBVRI}  & GDC $y(UBVRI)$\\
Table15 &{\sc DBA} &11000 K-40000 K & 7.0-9.0&  -2.0  &{u'g'r'i'z'y'} & LSM/Eq. 1 \\
Table16 &{\sc DBA} &11000 K-40000 K & 7.0-9.0&  -2.0  &{u'g'r'i'z'y'} & LSM/Eq. 2 \\
Table17 &{\sc DBA} &11000 K-40000 K & 7.0-9.0&  -2.0  &{u'g'r'i'z'y'} & LSM/Eq. 3 \\
Table18 &{\sc DBA} &11000 K-40000 K & 7.0-9.0&  -2.0  &{u'g'r'i'z'y'} & LSM/Eq. 4 \\
Table19 &{\sc DBA} &11000 K-40000 K & 7.0-9.0&  -2.0  &{u'g'r'i'z'y'} & LSM/Eq. 5 \\
Table20 &{\sc DBA} &11000 K-40000 K & 7.0-9.0&  -2.0  &{u'g'r'i'z'y'} & LSM/Eq. 6 \\
Table21 &{\sc DBA} &11000 K-40000 K & 7.0-9.0&  -2.0  &{u'g'r'i'z'y'} & GDC $y(u'g'r'i'z'y')$\\
Table22 &{\sc DBA} &11000 K-40000 K & 7.0-9.0&  -2.0  &{UBVRI}  & LSM/Eq. 1 \\
Table23 &{\sc DBA} &11000 K-40000 K & 7.0-9.0&  -2.0  &{UBVRI}  & LSM/Eq. 2 \\
Table24 &{\sc DBA} &11000 K-40000 K & 7.0-9.0&  -2.0  &{UBVRI}  & LSM/Eq. 3 \\
Table25 &{\sc DBA} &11000 K-40000 K & 7.0-9.0&  -2.0  &{UBVRI}  & LSM/Eq. 4 \\
Table26 &{\sc DBA} &11000 K-40000 K & 7.0-9.0&  -2.0  &{UBVRI}  & LSM/Eq. 5 \\
Table27 &{\sc DBA} &11000 K-40000 K & 7.0-9.0&  -2.0  &{UBVRI}  & LSM/Eq. 6 \\
Table28 &{\sc DBA} &11000 K-40000 K & 7.0-9.0&  -2.0  &{UBVRI}  & GDC $y(UBVRI)$\\
Table29 &{\sc DA}  &3750 K-60000 K & 5.0-9.5&    0.0 &{u'g'r'i'z'y'} & LSM/Eq. 1 \\
Table30 &{\sc DA}  &3750 K-60000 K & 5.0-9.5&    0.0 &{u'g'r'i'z'y'} & LSM/Eq. 2 \\
Table31 &{\sc DA}  &3750 K-60000 K & 5.0-9.5&    0.0 &{u'g'r'i'z'y'} & LSM/Eq. 3 \\
Table32 &{\sc DA}  &3750 K-60000 K & 5.0-9.5&    0.0 &{u'g'r'i'z'y'} & LSM/Eq. 4 \\
Table33 &{\sc DA}  &3750 K-60000 K & 5.0-9.5&    0.0 &{u'g'r'i'z'y'} & LSM/Eq. 5 \\
Table34 &{\sc DA}  &3750 K-60000 K & 5.0-9.5&    0.0 &{u'g'r'i'z'y'} & LSM/Eq. 6 \\
Table35 &{\sc DA}  &3750 K-60000 K & 5.0-9.5&    0.0 &{u'g'r'i'z'y'} & GDC $y(u'g'r'i'z'y')$\\
Table36 &{\sc DA}  &3750 K-60000 K & 5.0-9.5&    0.0 &{UBVRI}  & LSM/Eq. 1 \\
Table37 &{\sc DA}  &3750 K-60000 K & 5.0-9.5&    0.0 &{UBVRI}  & LSM/Eq. 2 \\
Table38 &{\sc DA}  &3750 K-60000 K & 5.0-9.5&    0.0 &{UBVRI}  & LSM/Eq. 3 \\
Table39 &{\sc DA}  &3750 K-60000 K & 5.0-9.5&    0.0 &{UBVRI}  & LSM/Eq. 4 \\
Table40 &{\sc DA}  &3750 K-60000 K & 5.0-9.5&    0.0 &{UBVRI}  & LSM/Eq. 5 \\
Table41 &{\sc DA}  &3750 K-60000 K & 5.0-9.5&    0.0 &{UBVRI}  & LSM/Eq. 6 \\
Table42 &{\sc DA}  &3750 K-60000 K & 5.0-9.0&    0.0 &{UBVRI}  & GDC $y(UBVRI)$\\
Table43 &{\sc DA-NLTE}  &40000 K-100000 K & 6.0-9.5&    0.0 &{u'g'r'i'z'y'} & LSM/Eq. 1 \\
Table44 &{\sc DA-NLTE}  &40000 K-100000 K & 6.0-9.5&    0.0 &{u'g'r'i'z'y'} & LSM/Eq. 2 \\
Table45 &{\sc DA-NLTE}  &40000 K-100000 K & 6.0-9.5&    0.0 &{u'g'r'i'z'y'} & LSM/Eq. 3 \\
Table46 &{\sc DA-NLTE}  &40000 K-100000 K & 6.0-9.5&    0.0 &{u'g'r'i'z'y'} & LSM/Eq. 4 \\
Table47 &{\sc DA-NLTE}  &40000 K-100000 K & 6.0-9.5&    0.0 &{u'g'r'i'z'y'} & LSM/Eq. 5 \\
Table48 &{\sc DA-NLTE}  &40000 K-100000 K & 6.0-9.5&    0.0 &{u'g'r'i'z'y'} & LSM/Eq. 6 \\
Table49 &{\sc DA-NLTE}  &40000 K-100000 K & 6.0-9.5&    0.0 &{u'g'r'i'z'y'} & GDC $y(u'g'r'i'z'y')$\\
Table50 &{\sc DA-NLTE}  &40000 K-100000 K & 6.0-9.5&    0.0 &{UBVRI}  & LSM/Eq. 1 \\
Table51 &{\sc DA-NLTE}  &40000 K-100000 K & 6.0-9.5&    0.0 &{UBVRI}  & LSM/Eq. 2 \\
Table52 &{\sc DA-NLTE}  &40000 K-100000 K & 6.0-9.5&    0.0 &{UBVRI}  & LSM/Eq. 3 \\
Table53 &{\sc DA-NLTE}  &40000 K-100000 K & 6.0-9.5&    0.0 &{UBVRI}  & LSM/Eq. 4 \\
Table54 &{\sc DA-NLTE}  &40000 K-100000 K & 6.0-9.5&    0.0 &{UBVRI}  & LSM/Eq. 5 \\
Table55 &{\sc DA-NLTE}  &40000 K-100000 K & 6.0-9.5&    0.0 &{UBVRI}  & LSM/Eq. 6 \\
Table56 &{\sc DA-NLTE}  &40000 K-100000 K & 6.0-9.0&    0.0 &{UBVRI}  & GDC $y(UBVRI)$\\
\hline
\hline
\end{tabular}
\end{flushleft}
\end{table*}
        \renewcommand{\tablename}{Table }       
\begin{table*}  
        \caption[]{Gravity and limb-darkening coefficients for the  HiPERCAM  photometric system}
        \begin{flushleft}
                \begin{tabular}{lcccccclc}                         
                        \hline                         
                        Name    & Source   &  range T$_{\rm eff}$ & range log $g$ & log [H/He] & Filters & Fit/equation   \\ 
                        \hline   
Table57  &{\sc DB}  &10000 K-40000 K & 5.5-9.5&  -10.0 &{HiPERCAM} & LSM/Eq. 1 \\
Table58  &{\sc DB}  &10000 K-40000 K & 5.5-9.5&  -10.0 &{HiPERCAM} & LSM/Eq. 2 \\
Table59  &{\sc DB}  &10000 K-40000 K & 5.5-9.5&  -10.0 &{HiPERCAM} & LSM/Eq. 3 \\
Table60  &{\sc DB}  &10000 K-40000 K & 5.5-9.5&  -10.0 &{HiPERCAM} & LSM/Eq. 4 \\
Table61  &{\sc DB}  &10000 K-40000 K & 5.5-9.5&  -10.0 &{HiPERCAM} & LSM/Eq. 5 \\
Table62  &{\sc DB}  &10000 K-40000 K & 5.5-9.5&  -10.0 &{HiPERCAM} & LSM/Eq. 6 \\
Table63  &{\sc DB}  &10000 K-40000 K & 5.5-9.5&  -10.0 &{HiPERCAM} & GDC $y(HiPERCAM)$\\
Table64  &{\sc DBA}  &11000 K-40000 K & 7.0-9.0& -2.0 &{HiPERCAM} & LSM/Eq. 1 \\
Table65  &{\sc DBA}  &11000 K-40000 K & 7.0-9.0& -2.0 &{HiPERCAM} & LSM/Eq. 2 \\
Table66  &{\sc DBA}  &11000 K-40000 K & 7.0-9.0& -2.0 &{HiPERCAM} & LSM/Eq. 3 \\
Table67  &{\sc DBA}  &11000 K-40000 K & 7.0-9.0& -2.0 &{HiPERCAM} & LSM/Eq. 4 \\
Table68  &{\sc DBA}  &11000 K-40000 K & 7.0-9.0& -2.0 &{HiPERCAM} & LSM/Eq. 5 \\
Table69  &{\sc DBA}  &11000 K-40000 K & 7.0-9.0& -2.0 &{HiPERCAM} & LSM/Eq. 6 \\
Table70  &{\sc DBA}  &11000 K-40000 K & 7.0-9.0& -2.0 &{HiPERCAM} & GDC $y(HiPERCAM)$\\
Table71  &{\sc DA}  &3750 K-60000 K & 5.0-9.0&   0.0 &{HiPERCAM} & LSM/Eq. 1 \\
Table72  &{\sc DA}  &3750 K-60000 K & 5.0-9.0&   0.0 &{HiPERCAM} & LSM/Eq. 2 \\
Table73  &{\sc DA}  &3750 K-60000 K & 5.0-9.0&   0.0 &{HiPERCAM} & LSM/Eq. 3 \\
Table74  &{\sc DA}  &3750 K-60000 K & 5.0-9.0&   0.0 &{HiPERCAM} & LSM/Eq. 4 \\
Table75  &{\sc DA}  &3750 K-60000 K & 5.0-9.0&   0.0 &{HiPERCAM} & LSM/Eq. 5 \\
Table76  &{\sc DA}  &3750 K-60000 K & 5.0-9.0&   0.0 &{HiPERCAM} & LSM/Eq. 6 \\
Table77  &{\sc DA}  &3750 K-60000 K & 5.0-9.0&   0.0 &{HiPERCAM} & GDC $y(HiPERCAM)$\\
Table78 &{\sc DA-NLTE}  &40000 K-100000 K & 6.0-9.5&    0.0 &{HiPERCAM} & LSM/Eq. 1 \\
Table79 &{\sc DA-NLTE}  &40000 K-100000 K & 6.0-9.5&    0.0 &{HiPERCAM} & LSM/Eq. 2 \\
Table80 &{\sc DA-NLTE}  &40000 K-100000 K & 6.0-9.5&    0.0 &{HiPERCAM} & LSM/Eq. 3 \\
Table81 &{\sc DA-NLTE}  &40000 K-100000 K & 6.0-9.5&    0.0 &{HiPERCAM} & LSM/Eq. 4 \\
Table82 &{\sc DA-NLTE}  &40000 K-100000 K & 6.0-9.5&    0.0 &{HiPERCAM} & LSM/Eq. 5 \\
Table83 &{\sc DA-NLTE}  &40000 K-100000 K & 6.0-9.5&    0.0 &{HiPERCAM} & LSM/Eq. 6 \\
Table84 &{\sc DA-NLTE}  &40000 K-100000 K & 6.0-9.5&    0.0 &{HiPERCAM} & GDC $y(HiPERCAM)$\\
\hline
\hline
\end{tabular}
\end{flushleft}
\end{table*}

        \renewcommand{\tablename}{Table }       
\begin{table*}  
        
        \caption{Gravity and limb-darkening coefficients for the Kepler, TESS, and Gaia  photometric systems}
        \begin{flushleft}
                \begin{tabular}{lcccccclc}                         
                        \hline                         
                        Name    & Source   &  range T$_{\rm eff}$ & range log $g$ & log [H/He] & Filters & Fit/equation   \\ 
                        \hline   
                        Table85 &{\sc DB}  &10000 K-40000 K & 5.5-9.5&  -10.0 &{KeplerTESSGaia} & LSM/Eq. 1 \\
                        Table86  &{\sc DB}  &10000 K-40000 K & 5.5-9.5&  -10.0 &{KeplerTESSGaia} & LSM/Eq. 2 \\
                        Table87  &{\sc DB}  &10000 K-40000 K & 5.5-9.5&  -10.0 &{KeplerTESSGaia} & LSM/Eq. 3 \\
                        Table88  &{\sc DB}  &10000 K-40000 K & 5.5-9.5&  -10.0 &{KeplerTESSGaia} & LSM/Eq. 4 \\
                        Table89  &{\sc DB}  &10000 K-40000 K & 5.5-9.5&  -10.0 &{KeplerTESSGaia} & LSM/Eq. 5 \\
                        Table90  &{\sc DB}  &10000 K-40000 K & 5.5-9.5&  -10.0 &{KeplerTESSGaia} & LSM/Eq. 6 \\
                        Table91  &{\sc DB}  &10000 K-40000 K & 5.5-9.5&  -10.0 &{KeplerTESSGaia} & GDC $y(KeplerTESSGaia)$\\
                        Table92  &{\sc DBA} &11000 K-40000 K & 7.0-9.0&  -2.0  &{KeplerTESSGaia} & LSM/Eq. 1 \\
                        Table93  &{\sc DBA} &11000 K-40000 K & 7.0-9.0&  -2.0  &{KeplerTESSGaia} & LSM/Eq. 2 \\
                        Table94  &{\sc DBA} &11000 K-40000 K & 7.0-9.0&  -2.0  &{KeplerTESSGaia} & LSM/Eq. 3 \\
                        Table95  &{\sc DBA} &11000 K-40000 K & 7.0-9.0&  -2.0  &{KeplerTESSGaia} & LSM/Eq. 4 \\
                        Table96  &{\sc DBA} &11000 K-40000 K & 7.0-9.0&  -2.0  &{KeplerTESSGaia} & LSM/Eq. 5 \\
                        Table97  &{\sc DBA} &11000 K-40000 K & 7.0-9.0&  -2.0  &{KeplerTESSGaia} & LSM/Eq. 6 \\
                        Table98  &{\sc DBA} &11000 K-40000 K & 7.0-9.0&  -2.0  &{KeplerTESSGaia} & GDC $y(KeplerTESSGaia)$\\
                        Table99  &{\sc DA}  &3750 K-60000 K & 5.0-9.5&    0.0 &{KeplerTESSGaia} & LSM/Eq. 1 \\
                        Table100 &{\sc DA}  &3750 K-60000 K & 5.0-9.5&    0.0 &{KeplerTESSGaia} & LSM/Eq. 2 \\
                        Table101 &{\sc DA}  &3750 K-60000 K & 5.0-9.5&    0.0 &{KeplerTESSGaia} & LSM/Eq. 3 \\
                        Table102 &{\sc DA}  &3750 K-60000 K & 5.0-9.5&    0.0 &{KeplerTESSGaia} & LSM/Eq. 4 \\
                        Table103 &{\sc DA}  &3750 K-60000 K & 5.0-9.5&    0.0 &{KeplerTESSGaia} & LSM/Eq. 5 \\
                        Table104 &{\sc DA}  &3750 K-60000 K & 5.0-9.5&    0.0 &{KeplerTESSGaia} & LSM/Eq. 6 \\
                        Table105 &{\sc DA}  &3750 K-60000 K & 5.0-9.5&    0.0 &{KeplerTESSGaia} & GDC $y(KeplerTESSGaia)$\\
                        Table106 &{\sc DA-NLTE}  &40000 K-100000 K & 6.0-9.5&   0.0 &{KeplerTESSGaia} & LSM/Eq. 1 \\
                        Table107 &{\sc DA-NLTE}  &40000 K-100000 K & 6.0-9.5&   0.0 &{KeplerTESSGaia} & LSM/Eq. 2 \\
                        Table108 &{\sc DA-NLTE}  &40000 K-100000 K & 6.0-9.5&   0.0 &{KeplerTESSGaia} & LSM/Eq. 3 \\
                        Table109 &{\sc DA-NLTE}  &40000 K-100000 K & 6.0-9.5&   0.0 &{KeplerTESSGaia} & LSM/Eq. 4 \\
                        Table110 &{\sc DA-NLTE}  &40000 K-100000 K & 6.0-9.5&   0.0 &{KeplerTESSGaia} & LSM/Eq. 5 \\
                        Table111 &{\sc DA-NLTE}  &40000 K-100000 K & 6.0-9.5&   0.0 &{KeplerTESSGaia} & LSM/Eq. 6 \\
                        Table112 &{\sc DA-NLTE}  &40000 K-100000 K & 6.0-9.5&   0.0 &{KeplerTESSGaia} & GDC $y(KeplerTESSGaia)$\\
                        \hline
                        \hline
                \end{tabular}
        \end{flushleft}
\end{table*}
\end{appendix}

\end{document}